\documentclass[letterpaper]{article}
\usepackage{aaai}

\usepackage{xspace}
\usepackage{graphicx}
\usepackage{algorithm}
\usepackage{algorithmic}
\usepackage{amsmath}
\usepackage{amssymb}

\usepackage{color}
\newcommand{\nop}[1]{{}}

\def\agg{\mathbf{agg}}

\usepackage{times}
\usepackage{helvet}
\usepackage{courier}
\begin{document}

\title{Mining for Spatially-Near Communities in Geo-Located Social Networks}
\author{
Joseph Hannigan\\
       {Dept.\ of Electrical Engineering}\\
			 {and Computer Science}\\
       {U.S.\ Military Academy}\\
       {West Point, NY}\\
       {Joseph.Hannigan@usma.edu}
\And
Guillermo Hernandez\\
       {Dept.\ of Electrical Engineering}\\
			 {and Computer Science}\\
       {U.S.\ Military Academy}\\
       {West Point, NY}\\
       {Guillermo.Hernandez@}\\
			 {usma.edu}
\And
Richard M. Medina\\
       {Dept.\ of Geography}\\
			 {and GeoInformation Science}\\
       {George Mason University}\\
       {Fairfax, VA}\\
       {rmedina3@gmu.edu}
\AND
Patrick Roos\\
       {Dept.\ of Computer Science}\\
       {University of Maryland}\\
       {College Park, MD}\\
       {roos@cs.umd.edu}
\And
Paulo Shakarian\\
       {Dept.\ of Electrical Engineering}\\
			 {and Computer Science}\\
       {U.S.\ Military Academy}\\
       {West Point, NY}\\
       {paulo@shakarian.net}
			}

\maketitle

\begin{abstract}
Current approaches to community detection in social networks often ignore the spatial location of the nodes.  In this paper, we look to extract spatially-near communities in a social network.  We introduce a new metric to measure the quality of a community partition in a geolocated social networks called ``spatially-near modularity'' a value that increases based on aspects of the network structure but decreases based on the distance between nodes in the communities.  We then look to find an optimal partition with respect to this measure - which should be an ``ideal'' community with respect to both social ties and geographic location. Though an NP-hard problem, we introduce two heuristic algorithms that attempt to maximize this measure and outperform non-geographic community finding by an order of magnitude.  Applications to counter-terrorism are also discussed.
\end{abstract}



\section{Introduction}
\label{sec:intro}

Community detection in social networks remains an important and active area of research in the study of social network mining \cite{girvan2002community,newman04,newman04a,du2007community,blondel08,Schaefer11,blondel11,cerina12,shak13kdd}.  However, many real-world social networks also have a geographic context.  Social networks are tethered to geographic locations. People and their relationships are tied to places. Even in the information age communications are dependent on access. Though access can seem ubiquitous in many cases, digital interaction cannot yet completely replace face-to-face contact, especially for planned activities of spatiotemporal coincidence and the transfer of tangible objects.

Primary considerations for research in many social science disciplines today include characteristics of human activities and interactions in defined spaces. The interactions can be between humans, or between humans and their environments. These characteristics describe aspects of social complexity that are necessary to understand when attempting to model open or closed social systems. In studies of human security there is a new emphasis on implementations of Activity Based Intelligence (ABI) to better understand drivers toward specific actions and interactions, as well as to generate an understanding of the system outside of targeted activities ~\cite{miller13}. Though the concepts of ABI are new to many, its academic foundations of activity spaces, social interaction, and spatio-temporal research are well established. 

Attempts to identify sociogeographic based activity spaces, as demonstrated here, are vital to the understanding of human behavior. Multi-spatial or hybrid space~\cite{batty00} studies are much more valuable in this information age than their single space counterparts. Multiple spaces are converging into hybrid spaces as interactions in social systems become more complex.

In this paper, we look to develop a framework for deriving communities from social network that is relevant not only with respect to network topology, but also geography. The main geographic concept we use to relate nodes based on space is ``nearness.''  On a general level, there exists a connection between nearness and similarity. ``Near'' is a spatial concept, though not necessarily geographically spatial. Social space nearness, or adjacency, typically describes relationships between people or things that interact in some way. Nearness based similarities need not be comprehensive. Single or few similar traits can exist to maintain interaction; however, relatively more similar traits between people can drive further or deeper interaction.  Geographers and sociologists have developed concepts that seek to explain the phenomenon of nearness and similarity in their respective disciplines. In geography Tobler's First Law (TFL) describes this effect in physical space and homophily describes it in social space~\cite{tobler70,mcpherson01}. Geographic and homophilic similarities are inherently connected, as one of the greatest sources of homophily is propinquity. Furthermore, interaction is driven by nearness and similarity. The likelihood for interaction between people increases as distance decreases between them. Community finding at the convergence of geographic and social space nearness will lead to the identification of communities where place and social traits drive interaction. Results of this method may be most meaningful in studies of social systems that are greatly influenced by ethnicity and culture among other geographically based factors. For example, communities identified using geographic and social closeness may apply more to terrorist and criminal networks than globally dispersed business networks.

Hence, our intuition is to find communities that are tightly-knit based on network topology, but also spatially ``near.''  To do this, we create a new measure of partition quality that we term ``spatial nearness modularity'' that borrows concepts from network modularity~\cite{newman04} and $K$-means clustering~\cite{macqueen1967some,lloyd1982least}.  Hence, to find a high-quality set of communities with respect to this geography and network connections, it stands to reason to search for an optimal partition with respect to this measure.  Unfortunately, we are able to show that doing so is NP-hard based on the results of \cite{brandes08}.  To address this issue of intractability, we introduce two heuristics and we then experimentally evaluate them, where we find that our approach provides an order-of-magnitude improvement in spatially-near modularity over non-geographic approaches.  This is followed by a description of how this technique could apply to counter-terrorism and a discussion of related work.

\section{Technical Preliminaries}
\label{prelimsec}

We assume the existence of a undirected graph $G=(V,E)$ where set $V$ are vertices and $E$ are edges among them.  As the graph in undirected, $(i,j) \in E$ implies $(j,i)\in E$.  We shall use $n,m$ to represent the sizes of $V,E$ respectively.  Each edge $(i,j)$ will be associated with a positive real weight denoted by $w_{ij}$ (if there is no edge between $i$ and $j$, $w_{ij}=0$).  For a given node $i \in V$, we shall use the symbol $\eta_i$ to represent the set $\{ j\in V | \exists (i,j) \in E\}$ and $k_i$ is the size of this set.  We shall also assume a distance function $d : V \times V \rightarrow \Re$ that meets the normal distance axioms: $d(i,i)=0$, $d(i,j)=d(j,i)$, and $d(i,j) \leq d(i,j')+d(j,j')$.  For ease of notation, we shall use $d_{ij}$ instead of $d(i,j)$.  In this paper we will often use the notation $C = c_1,\ldots,c_x$ to denote a partition of $V$.  Hence, $\cup_{c_i \in C} c_i = V$ and for all $c_i,c_j \in C$ $c_i \cap c_j = \emptyset$.  We define the \textit{modularity} of a partition ($M(C)$) in accordance with the definition introduced by \cite{newman04} as follows:\\

\noindent\textbf{NG-Modularity.~\cite{newman04}}  \textit{
Given a social network $G=(V,E)$ and partition $C$ the \textbf{Newman-Girvan (NG) modularity} is defined as follows:
\begin{eqnarray}
M_{NG}(C) = \frac{1}{2m}\sum_{c \in C} \sum_{i,j \in c} w_{ij}-\frac{k_i k_j}{2m}.
\end{eqnarray}}

The modularity of a network partition measures the quality of its partition structure as the density of edges within partitions compared to the density of edges between partitions. The former is ideally very high compared to the latter.  Modularity will give a number in $[-1,1]$, a higher value meaning better quality partition.  Previous work, such as \cite{brandes08,blondel11}, has focused on finding a partition that optimizes this quantity.  However, modularity maximization only considers network topology and does not make any effort to group individuals that are geographically close to each other.  An alternative is to find a partition of $K$ clusters of nodes that minimizes the sum-of-squares distance to the center of each cluster.  This is known as $K$-means clustering~\cite{macqueen1967some,lloyd1982least}. $K$-means clustering algorithms attempt to find a partition of points on a plane into $K$ clusters, such that the following quantity is minimized (here $x_c$ is the centroid of the points in cluster $c$):

\begin{eqnarray}
\sum_{c \in C}\agg_{i \in c}d(i,x_c)^2.
\end{eqnarray}

In the above definition, $\agg$ is some aggregate function.  Common aggregates used here are $\max$ and $\sum$.  For the purpose of this paper, as modularity is maximized, we wish to minimize some aggregate of the distances to the center of each cluster. Thus, one potential quantity that could be optimized is the following:

\begin{eqnarray}
\frac{1}{2m}\big(\frac{\sum_{c \in C} \sum_{i,j \in c} w_{ij}-\frac{k_i k_j}{2m}}{1+\sum_{c \in C}\agg_{i \in c}d(i,x_c)^2}\big).
\end{eqnarray}

Note that the additive $1$ in the denominator is to avoid division by zero and to ensure that the result will be within the range $[-1,1]$.  The above optimization function has the useful property that we can embed both modularity maximization and $K$-means clustering - the first by placing all nodes in the same location, the second by ignoring edges among any nodes in the network and restricting the number of clusters to be exactly $K$.  However, one aspect the above definition misses is that it cannot measure the quality of an individual community.  Hence, we introduce an alternative definition below that we term ``spatially-near (SN) modularity.''\\

\noindent\textbf{SN-Modularity.} \textit{
Given a social network $G=(V,E)$, partition $C$, and scaling parameter $\sigma \in \Re^+$, the \textbf{spatially-near (SN) modularity} is defined as follows:
\begin{eqnarray}
\label{optFcn}
M_{SN}(C,\sigma)=\frac{1}{2m}\sum_{c \in C}\frac{\sum_{i,j \in c} w_{ij}-\frac{k_i k_j}{2m}}{1+\agg_{i \in c}\left(\frac{d(i,x_c)}{\sigma}\right)^2}.
\end{eqnarray}}

So, for a given community, we can measure its quality with the following:

\begin{eqnarray}
\label{eaComm}
\frac{1}{2m}\frac{\sum_{i,j \in c} w_{ij}-\frac{k_i k_j}{2m}}{1+\agg_{i \in c}\left(\frac{d(i,x_c)}{\sigma}\right)^2}.
\end{eqnarray}

We also note that as $\sigma$ increases, distance is de-emphasized.  This parameter would be specified based on the relative importance of distance to to network structure as well as the unit of measurement used for distance.  Practically, a user could potentially provide this parameter in many different ways.  Simple methods would include setting $\sigma$ to $1$, the average distance among all pairs of nodes, or the average distance among all pairs of nodes that have an edge between them.  Alternatively, this parameter could also learned from historical data, if such a corpus is available.  Another approach is for the user to explore various parameter settings.  In this work, we leave advanced methods for determining $\sigma$ to future work and conduct experiments with multiple settings for this parameter.  However, we note that for particularly large values of $\sigma$, SN-modularity becomes equivalent to NG-modularity.  It is easy to show the following property:

\begin{eqnarray}
\label{limExpr}
\mathit{lim}_{\sigma \rightarrow \infty}M_{SN}(C,\sigma) = M_{NG}(C).
\end{eqnarray}

However, maximizing $M_{SN}(C,\sigma)$ remains NP-hard.  Hence, in this paper we introduce two heuristic algorithms to find a partition $C$ where $M_{SN}(C)$ is near-optimal.

\noindent\textbf{Theorem.} \textit{For a given social network $G=(V,E)$ and scaling factor $\sigma$, identifying a partition $C$ s.t. $M_{SN}(C,\sigma)$ is maximized is NP-hard.}\\
\noindent\textit{Proof.} We can embed an instance of finding a partition $C$ that maximizes $M_{NG}(C)$ into the problem from the statement by creating a distance function $d$ where $\forall i,j,  d(i,j)=0$ and setting $\sigma$ to an arbitrary value.  Hence, any algorithm that maximizes $M_{SN}$ using this construction also maximizes $M_{NG}$. Since finding a partition that maximizes $M_{NG}$ is NP-hard by the results of  \cite{brandes08}, the statement of the theorem follows.  $\blacksquare$\\
\section{Algorithms}
\label{sec:algs}

In the previous section, we found that identifying a spatially-near partition is an NP-hard problem.  Hence, in this section, we propose two heuristic approaches to deal with this intractability.  We later describe our evaluation of these approaches.  In our first heuristic, which we call ``Louvain-SN'', we employ the modification of the Louvain algorithm of Blondel et al.~\cite{blondel08}, only instead of using it to maximize NG-modularity, we use it to maximize SN-modularity (the Louvain algorithm was designed to find a near-optimal parition w.r.t. NG-modularity).  Our second algorithm, the SNIC (Spatially Near Iterative Constraining) algorithm, relies on multiple calls to the Louvain-SN algorithm - but each with a limit on the aggregate distance permitted in a community.

\subsection{The Louvain-SN Algorithm}
The original Louvain algorithm of \cite{blondel08} is an iterated, hierarchical process in which two phases are applied repeatedly until maximal modularity is reached: During the first phase, each node $v_i \in V$ of the given social network is assigned to a community $c$, creating an \textit{initial partition}. In~\cite{blondel08}, the singleton partition was used - which we use in this work as well. Then, for each $v_i \in V$, the gains in modularity that would result from moving $v_i$ to the community of each of its neighbors $v_j \in \eta_i$ are calculated, and $v_i$ is removed and placed into the community for which the maximum improvement in modularity is achieved (unless no positive gain in modularity is possible). This sub-process is repeated sequentially for each $v_i \in V$ until no individual move will result in a  gain in modularity, marking the end of the first phase and giving a partition $C$. 
During the second phase, a new network is built by using each $c_i \in C$ as a node in the new network, call these nodes \textit{meta-nodes}. Weights on the edges between any two meta-nodes in the new network are assigned to be the sum of the weights of the edges between nodes in the two communities corresponding to the meta-nodes. Here, self-loops are created for each meta-node in the new network from the links between nodes of the community corresponding to that meta-node. After this phase is complete, the two phases are reapplied iteratively until there are no more changes. 

The efficiency of the Louvain algorithm relies on an easy re-calculation of modularity in the first phase of the algorithm.  When computing gains in modularity in phase one of the algorithm, removing any node $v_i$, the overall increase in modularity if it is placed into community $c$ is proportional to:

\begin{eqnarray}
\label{incrEqn}
k_{i,in}-\sum_{j \in c}\frac{k_i k_j}{2m}
\end{eqnarray}

In our modification for optimizing SN-modularity, we can retain some of this efficiency by retaining the previous denominator and numerator of Equation~\ref{eaComm} (multiplied by $2m$) for each community.  By retaining these values along with the value of Equation~\ref{incrEqn}, computing the increase or decrease in modularity for a community can be performed quickly (though this ultimately depends on how the aggregate function $\agg$ and the centers of the communities $x_c$ are computed).  Additionally, in the creation of the meta-nodes, we use the centers from the previous step as their location.  Additionally, we also found that we obtained improvement in performance by allowing a removed node to be moved back to a community containing just itself, as unlike in the maximization of standard modularity, isolating a node in this fashion could potentially increase the overall modularity due to the denominator of Expression~\ref{optFcn}.

\subsection{The SNIC Heuristic}

Next, we introduce the ``Spatially Near, Iterative Constraining'' (SNIC) Heuristic.  This idea was created as the result from a pilot experiment where we noticed that constraining a node to join only communities where it was geographically ``near'' to all the members would sometimes improve the resulting quantity of $M_{SN}$.  The question is how does one determine where to set this distance constraint.  In our experiments we ran our modified Louvain approach iteratively, returning only the maximum distance between two points in the community upon each iteration.  This distance is then set as the distance constraint for the next iteration.  Once the distance constraint reaches zero (or a maximum number of iterations is reached), the algorithm then returns the partition found which is associated with the greatest value for Expression~\ref{optFcn}.

\section{Experimental Results}
\label{expsec}

For our experiments, we used information extracted from the Brightkite location-based online social networking sites~\cite{cho11}.  

We built our implementation in Python 2.6 on top of the NetworkX library\footnote{http://networkx.github.com/} leveraging code from Thomas Aynaud's implementation of the Louvain algorithm\footnote{http://perso.crans.org/aynaud/communities/}.  Our implementation took approximately $1000$ lines of code.    The experiments were run on a computer equipped with an Intel Core i7 Processor operating at 2.67 GHz (one core utilized) running Microsoft Windows 7 and equipped with 4.0 GB of physical memory.  All statistics presented in this section were calculated using SPSS 19.  We use our heuristics to find partitions based on Expression~\ref{optFcn} where $\agg=\max$.


In our first set of tests, we iteratively selected nodes and their neighbors from the Brightkite network dataset provided by the authors of \cite{cho11} to produce 10 samples of 1000 nodes. To generate the samples, each sample begins with a randomly selected node from the network. The selected node and all of its connected nodes are then included in the next iteration, in which a new random node is chosen. This continues until 1000 nodes are reached for each sample. The minimum edge count for all samples processed is 1729, while the maximum is 2282. The average number of edges is 1929.

In our trials, we varied the $\sigma$ parameter with the values $\{300,500,1000,2000,3000,4000,5000\}$.  For each dataset and each value of $\sigma$, we compare the SN-modularity returned by three approaches: the Louvain algorithm (does not consider geospatial information), the Louvain-SN algorithm (the modified version of the Louvain algorithm for SN-modularity optimization), and the SNIC algorithm (an iterated version of the Louvain-SN that selects the best result based on updating the distance constraint). 

\begin{figure}[h]
    \begin{center}
        \includegraphics[width=.8\linewidth]{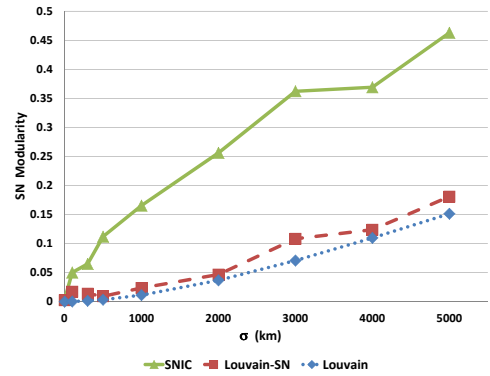}
    \end{center}
    \caption{$\sigma$ (in kilometers) vs. (average) SN-modularity for the partitions returned by the Louvain, Louvain-SN, and SNIC algorithms.}
    \label{fig:avg_SN_Mod}
\end{figure}

The SNIC algorithm returned a partition with greater average SN-modularity for each value of $\sigma$ than the partitions returned by the Louvain and Louvain-SN algorithms (see Figure ~\ref{fig:avg_SN_Mod}).  In general, the SNIC algorithm consistently outperformed the Louvain algorithm in terms of SN-modularity - producing a partition of greater SN-modularity on all trials.  The Louvain-SN outperformed the standard Louvain in all but $11$ (of $70$) trials, though (as we discuss later in this section) this improvement is likely not statistically significant, unlike the SNIC heuristic.\footnote{There were $11$ such trials out of the $70$ trials where the Louvain outperformed the Louvain-SN.  Of the cases where there was decreased quality over standard Louvain, the maximum decrease in quality was $26.52\%$ and the average decrease was $15.10\%$.  The SNIC algorithm outperformed the Louvain algorithm on all trials.}

\begin{figure}[h]
    \begin{center}
        \includegraphics[width=.8\linewidth]{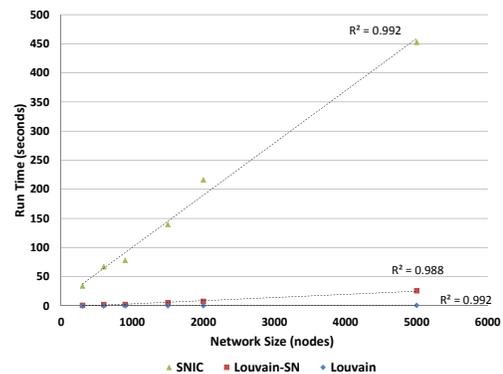}
    \end{center}
    \caption{Network size (by nodes) vs. runtime for the partitions returned by the Louvain, Louvain-SN, and SNIC algorithms.}
    \label{fig:NetSize_Runtime}
\end{figure}

To determine significant difference in SN-modularity of the three approaches on the Brightkite dataset, analysis of variance (ANOVA) tests were used. Difference in SN-modularity for the three approaches was confirmed with a p-value of 0.006. A Tukey's Honest Significant Difference (HSD) test was also used to determine pairwise differences between the approaches. No significant differences were found between the Louvain and Louvain-SN algorithms; however, the SNIC algorithm was found to be different than both the Louvain (at p = 0.010) and the Louvain-SN (at p = 0.020). Additionally, the differences for runtimes of the three approaches were found to be different with a p-value of 0.000 (see Figure ~\ref{fig:NetSize_Runtime}). As with the difference in SN-modularity, runtime differences exist between the Louvain and the SNIC algorithms (at p = 0.000) and between the Louvain-SN and the SNIC algorithms (at p = 0.000). These results are also provided through use of Tukey's HSD. Differences in SN-modularity and runtimes for the three approaches can be seen in Figures ~\ref{fig:avg_SN_Mod} and ~\ref{fig:NetSize_Runtime}, respectively.  Further, we also note that although the SNIC algorithm has significantly greater runtime than the Louvain and Louvain-SN, it still appear to scale linearly with the number of nodes in the network ($R^2=0.992$).  Hence, it may still be a viable solution for very large networks.  We are currently studying the scalability of this algorithm.

\begin{figure}[h]
    \begin{center}
        \includegraphics[width=.8\linewidth]{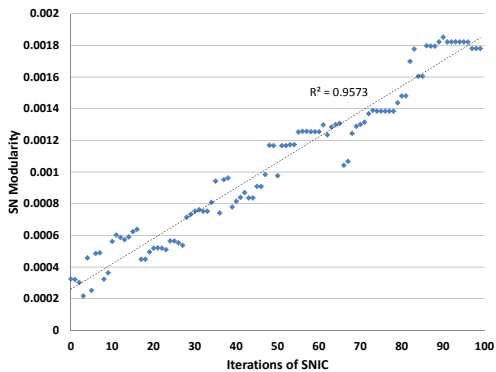}
    \end{center}
    \caption{SN-modularity vs. number of iterations for the partitions returned by the SNIC algorithms for the Britekite network data.}
    \label{fig:Mod_Iter}
\end{figure}

Figure ~\ref{fig:Mod_Iter} shows the increase in quality of community finding (SN-modularity) over iterations of the SNIC algorithm. Recall that the SNIC algorithm decreases the distance constraint at each iteration. As the geographic constraint decreases, such that community proximity becomes more important, the quality of community (number of connections within vs outside) increases. Here we introduce an axiom - that as the geographic space of interaction for a social network shrinks, it is more likely that those left within the community are more connected. Spatial outliers, which are also social outliers can be conceptualized as weak links~\cite{gran} and are removed through community proximity limiting iterations. Through 100 iterations, the quality of community increases and in most social networks this value may continue to increase given high enough spatial resolution data. In other words, humans form communities and interact mostly with those they are geographically near, such that the strongest communities will be those shared within small geographic proximities.  However, we note that more iterations of the SNIC algorithm will not result in singleton communities, as that is the initial partition considered by the algorithm.  Also note that the improvement in SN-modularity as a function of number of iterations of the SNIC heuristic is also likely dependent on the parameter $\sigma$.

\begin{figure}[h]
    \begin{center}
        \includegraphics[width=.8\linewidth]{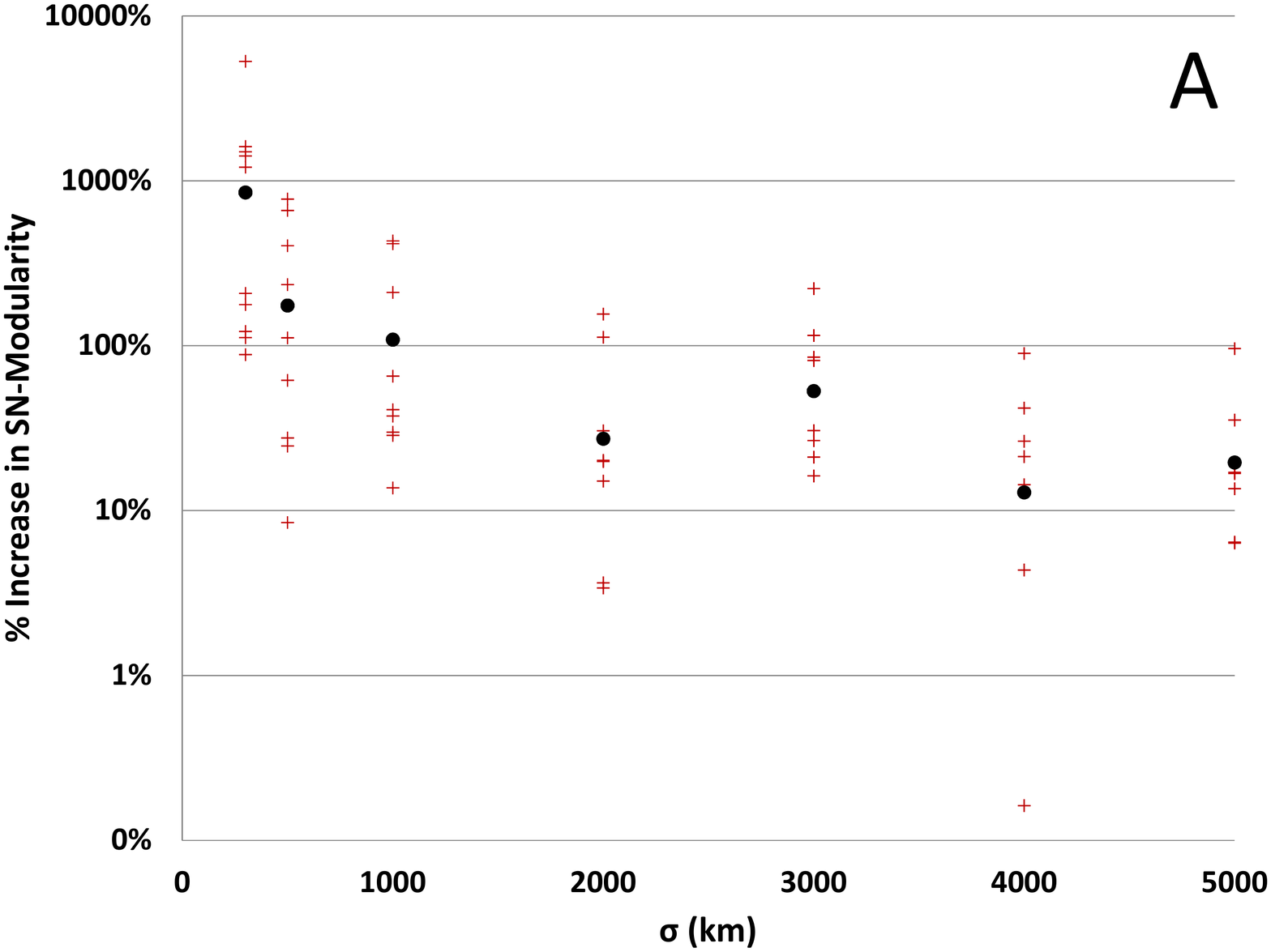}
				\includegraphics[width=.8\linewidth]{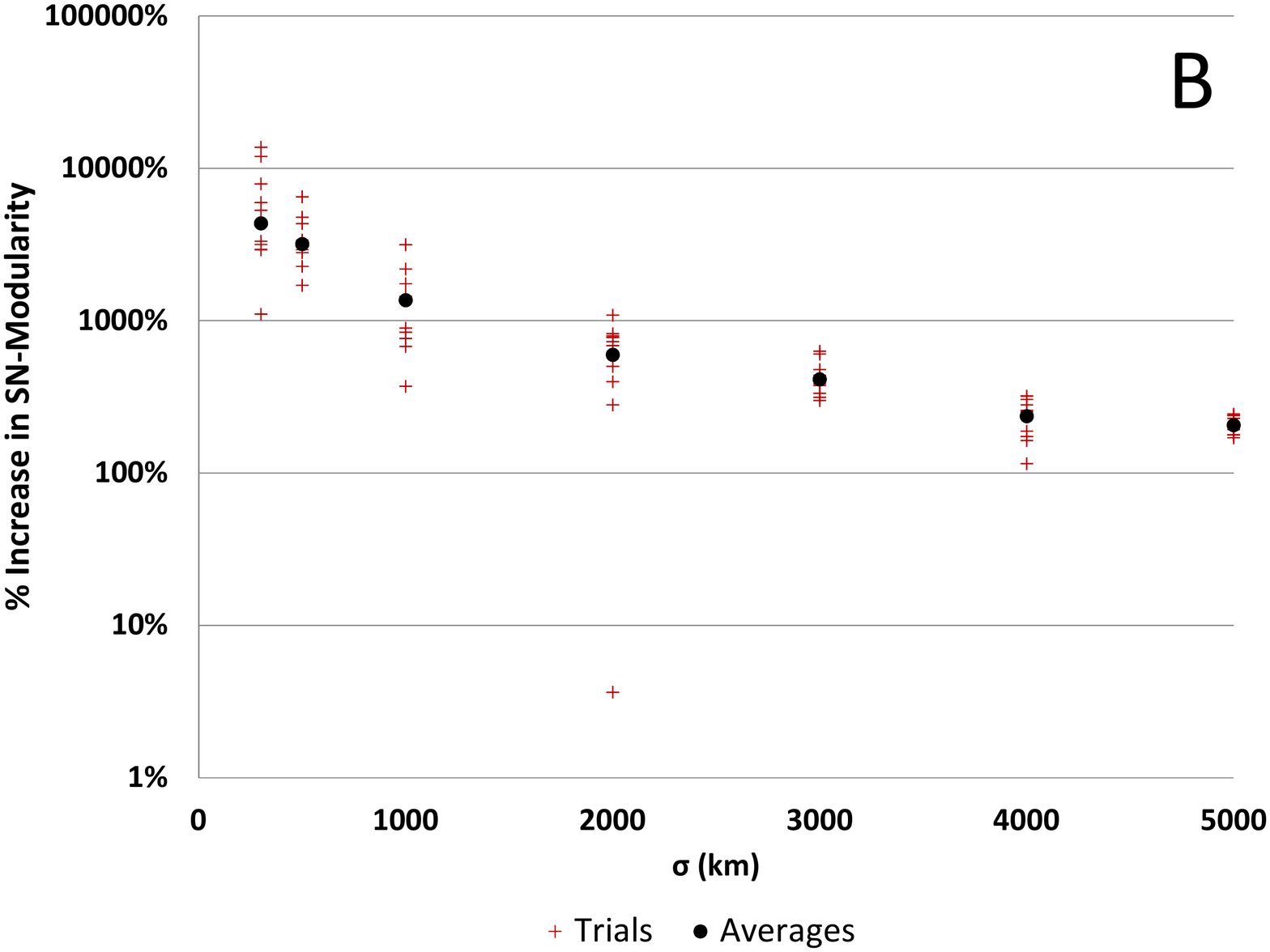}
    \end{center}
    \caption{$\sigma$ (in kilometers) vs. percent improvement in SN-modularity for the partition returned by the Louvain-SN (panel \textbf{A}) and SNIC (panel \textbf{B}) algorithms.  Not depicted in panel \textbf{A} (Louvain-SN) are results where the Louvain-SN algorithm produced lower-quality results than the standard Louvain (due to the log-scale, see text for further details).$^3$  Note that for the SNIC algorithm (panel \textbf{B}) outperformed the standard Louvain on all trials.}
    \label{fig:Inc_L_L-SN}
\end{figure}

As with the example above in Figure ~\ref{fig:Mod_Iter}, Figure ~\ref{fig:Inc_L_L-SN} shows that a stronger influence of geographic distance on community finding leads to greater quality communities based on the SN-modularity measure. Recall that $\sigma$ is the scaling parameter in SN-modularity. Decreasing the scaling parameter, in turn strengthening the geographic influence on the equation, leads to an increase in the quality of communities identified by the SNIC algorithm. Increasing the $\sigma$ value will result in an asymptotic trend for SN-modularity toward that expected from the non-spatial Louvain algorithm. This trend is shown for the Brightkite network in percent increase in SN-modularity over the Louvain algorithm for (A) the Louvain-SN and  (B) the SNIC algorithms.

\begin{figure}[h]
    \begin{center}
        \includegraphics[width=.8\linewidth]{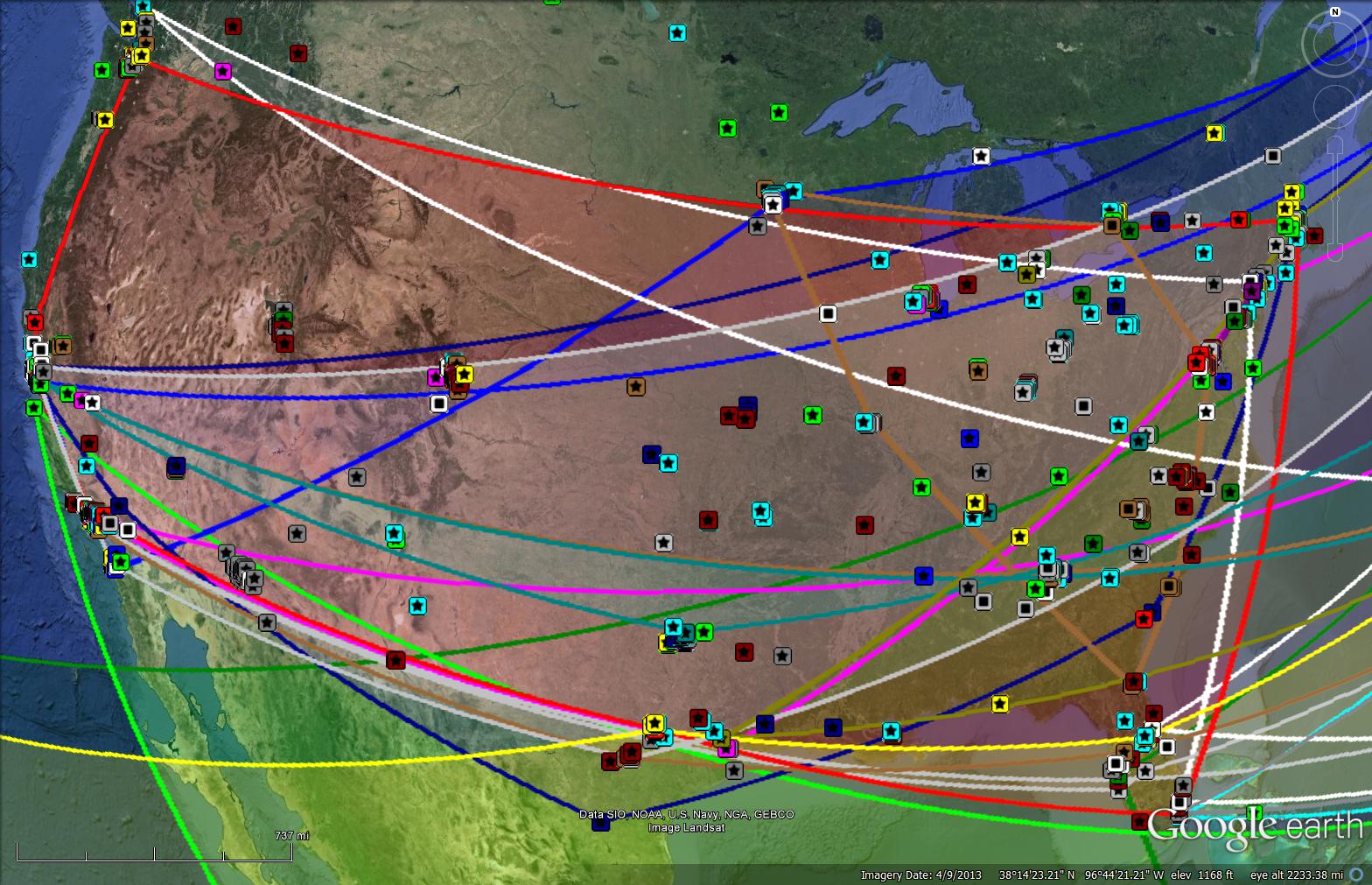}
				
				\vspace{1 mm}
			
				\includegraphics[width=.8\linewidth]{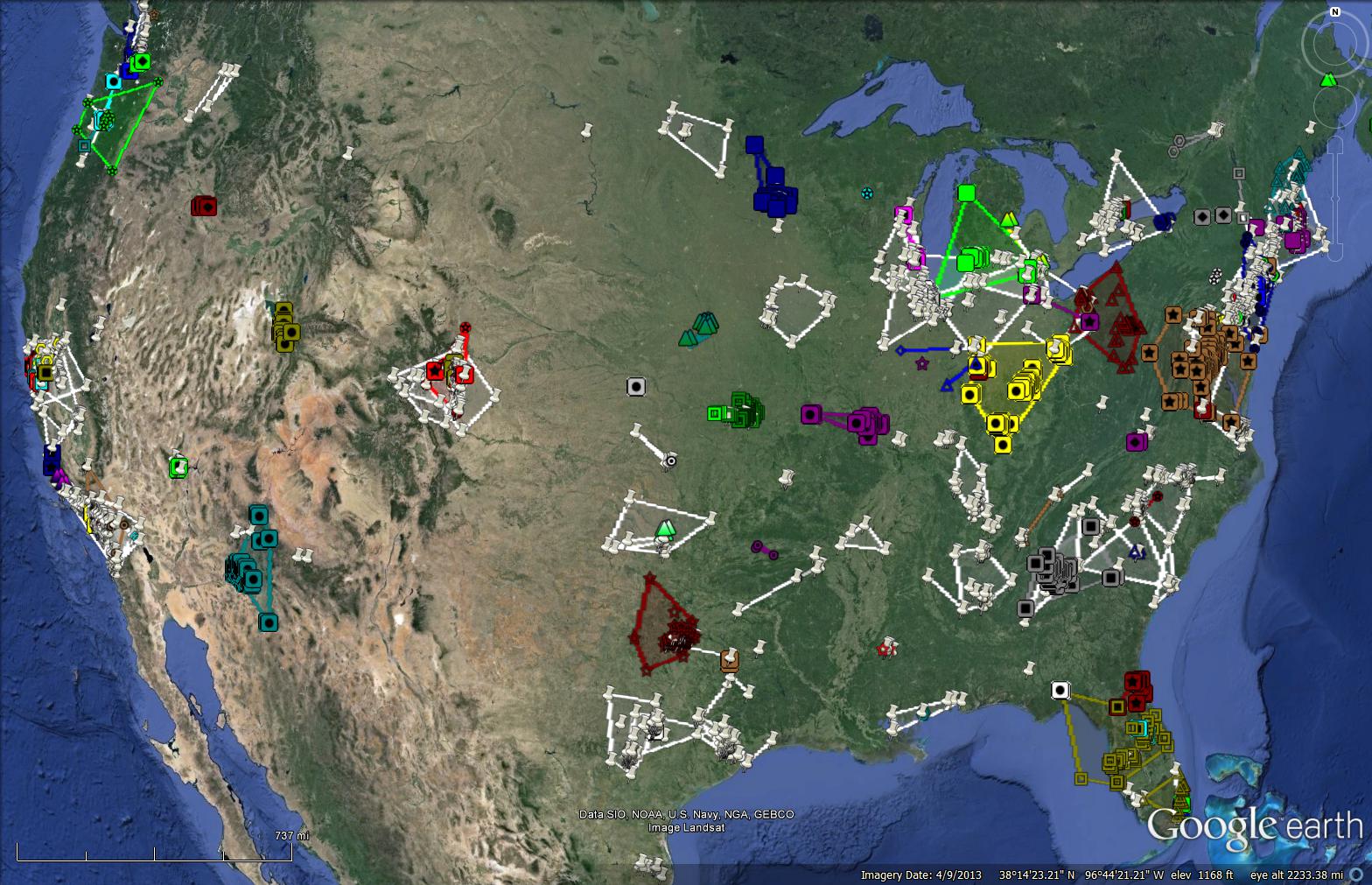}
    \end{center}
    \caption{Top: Brightkite Communities identified using the Louvain algorithm, Bottom: Communities identified using the SNIC algorithm.}
    \label{fig:BK_Louvain}
\end{figure}

The difference in Brightkite communities identified by the Louvain and SNIC algorithms is clear in Figure ~\ref{fig:BK_Louvain}. There is a quantitative difference as suggested by the SN-modularity metric results, but also a very qualitative difference in which the communities identified by the SNIC algorithm are much more spatially constrained. The bottom half of Figure ~\ref{fig:BK_Louvain} represents the SNIC algorithm results with $\sigma$ = 1. In today's information age where global networks are common, methods to identify geograhically unconstrained communities, as well as those methods that identify their geographically constrained counterparts are both equally valuable. Implications for strength of ties, activity and operations spaces, and interactions are different when considering geographic network characteristics.

\begin{figure}[h]
    \begin{center}
        \includegraphics[width=.8\linewidth]{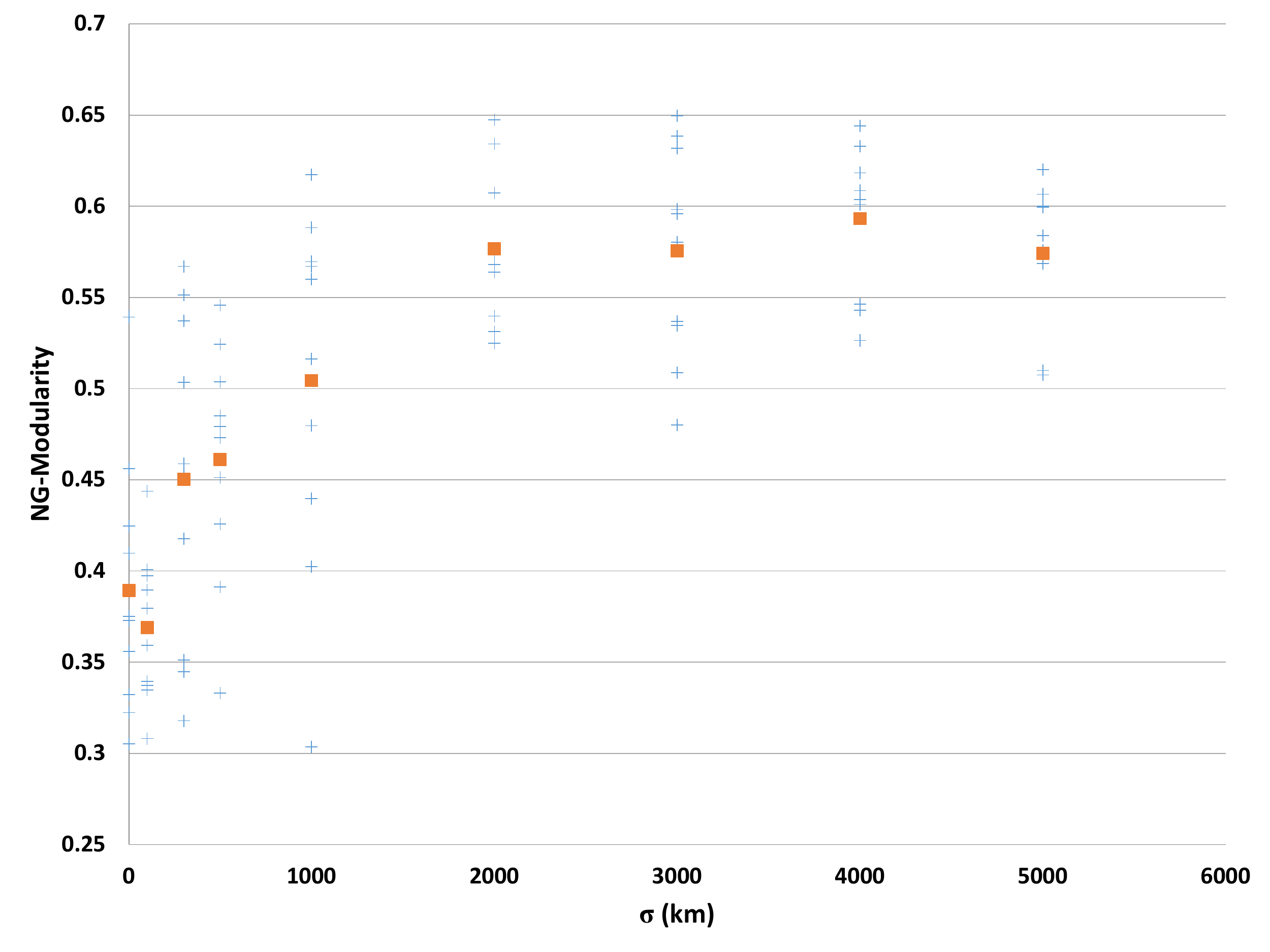}
    \end{center}
    \caption{NG-modularity of the partition returned by the SNIC algorithm ($10$ iterations) as a function of $\sigma$.}
    \label{fig:ngSnic}
\end{figure}

Additionally, we also studied the NG-modularity of the partition returned by the SNIC algorithm.  We found that although the SNIC algorithm was not designed to maximize NG-modularity, it still provided a positive value - which indicates that there is a greater density of edges within the communities as opposed to between communities (Figure~\ref{fig:ngSnic}).  We also found that the solution returned by the NG-modularity of the partition returned by the SNIC algorithm seems to approach the NG-modularity of the solution of the Louvain algorithm as $\sigma$ increases.  Although this is not guaranteed theoretically, it should be expected based on the relationship between NG-modularity and SN-modularity shown in Equation~\ref{limExpr}.

\section{Applications}
\label{apps}
There are many fitting applications for algorithms that detect sociogeographic communities. In general, any network that requires or benefits from geographic propinquity can serve as a test case for the SNIC algorithm. For example, the diffusion of a disease through a social network requires geographic closeness, or face-to-face interaction, between people. While much of the diffusion may not be social network based, but solely spatial, those that have stronger social ties and in turn interact more in geographic space are more likely to contract or spread a disease. This phenomenon exists at various levels of physical interaction for contagious diseases. This model of diffusion works with the spread of any biologically contagious, material, or even ideological transfer that requires coincidence in space and time. The following example shows the value of the SNIC algorithm for sociogeographic analysis on a transnational terrorist network.  

\begin{figure}[h]
    \begin{center}
        \includegraphics[width=.8\linewidth]{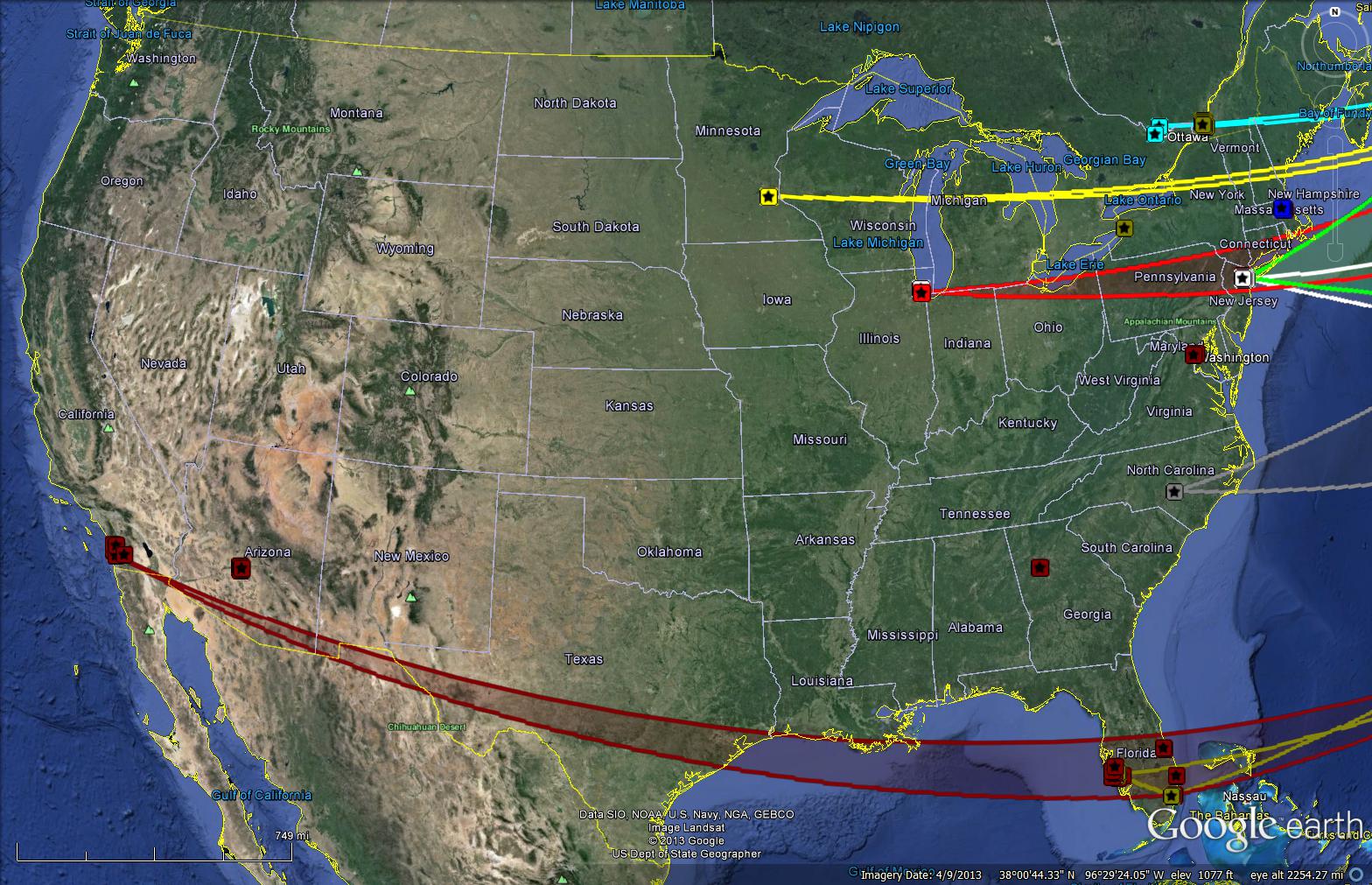}
				
				\vspace{1 mm}
				
				\includegraphics[width=.8\linewidth]{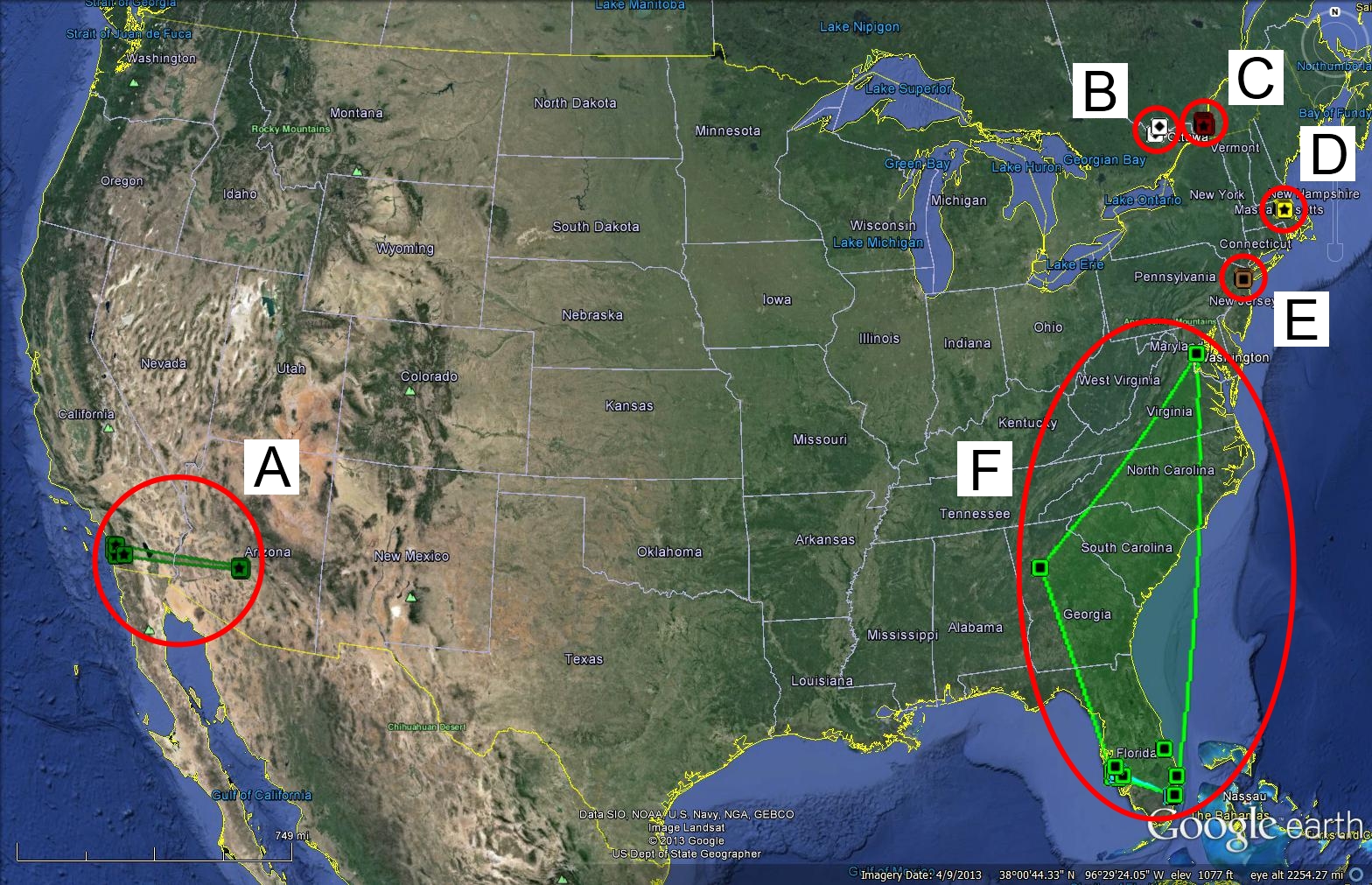}
    \end{center}
    \caption{Top: Terrorist communities identified using the Louvain algorithm, Bottom: Terrorist communities identified using the SNIC algorithm.}
    \label{fig:TNET}
\end{figure}

Figure~\ref{fig:TNET} illustrates the difference between community finding results using both the Louvain (non-spatial) and SNIC algorithms on a transnational terrorist network dataset. The dataset used for this research is representative of a global Islamist terrorist network from the late 1970s to approximately 2010 (see ~\cite{medina} for a full description and discussion of the dataset). The SNIC algorithm application shown here uses $\sigma$ = 1600. The transnational Islamist terrorist network is a cellular based, decentralized structure and heavily dependent on relative location and proximity ~\cite{MH2013}. Because this is the case, identifying sociogeographic communities requires only a small spatial component additional to the social component. Applications of the SNIC algorithm on other network structures may require more spatial influence to identify sociogeographic communities. For example, the Brightkite application shown in Figure~\ref{fig:BK_Louvain} uses $\sigma$ = 1. The terrorist network is much smaller with 358 nodes and 660 edges, and is much more geographically based for operational necessity.

As stated previously, research results that identify social closeness vs. those that identify sociogeographic closeness are quantitatively and qualitatively different for many social networks. The top graphic in Figure~\ref{fig:TNET} shows the modularity results using the Louvain algorithm, which highlight the transnational network connections. Many of the Islamist terrorist network cells have foundations or affiliates in Europe and the Middle East. While the strength of social communities can be equal over long distances, especially if network connections were made at some point coincident in space and time, it is beneficial to isolate communities in geographic space for some applications. In this case, the SNIC algorithm successfully identifies operational communities (A) the 9/11 cell planning and preparing for the attack in Southern California and Arizona, (B) a father and son diad working with, specifically financing, al-Qaeda in Canada, (C) a sociogeographic community of al-Qaeda tied members in Montreal, Canada, some of which were plotting to attack Los Angeles International Airport in 1999, (D) two al-Qaeda linked cells in Boston, MA with members in the Boston sleeper cell and plotting a large scale bombing attack in Jordan at multiple sites, (E) the al-Qaeda based cell operating in New York responsible for the first World Trade Center Attack in 1993, and (F) communities of 9/11 hijackers operating in Florida and other eastern US states. The SNIC algorithm can be additionally adjusted to further separate cellular communities based on geography (by varying $\sigma$ and the number of iterations of the algorithm). 

In systems such as this terrorist network, connected individuals that are close in geographic space, but not as close socially, can be more important to identify when attempting to counter operations. For example, identification of weaker but closer social links, such as those providing materials to a terrorist cell can be used as valuable intelligence to understand and dismantle terrorist operations in local to regional settings. Knowledge of international connections is important for understanding the global terrorist system, and cells in decentralized networks often maintain communications over long distances. However, many of these cells can operate independently, though in most cases they will need proximal resources. These local system interactions can be detected through use of the SNIC algorithm.

\section{Related Work}
\label{rw}

Modularity maximization for community finding was first introduced in \cite{newman04}. In \cite{blondel08}, the Louvain algorithm is introduced, which can scale to very large networks and is shown to provide partitions that nearly maximize modularity. We leverage a modification of the Louvain algorithm in this paper.  Finding geographically disperse communities in a social network has also been previously studied~\cite{shak13kdd,liu12,cerina12,blondel11}.  Our approach in this paper differs in that we desire to find communities where the nodes are spatially-near and not distant.  In addition to the aforementioned approaches, community detection in networks has also been explored in other manners that have potential  to be applicable to the geospatial case - though to our knowledge no such application has been presented in the literature.  See \cite{yang09,mucha2010community} for examples.

There also exist many approaches for community detection in networks not based on modularity maximization. Examples use label propagation \cite{raghavan2007near}, random walks \cite{rosvall2008maps}, or bottom-up voting approaches \cite{DBLP:conf/kdd/CosciaRGP12}. See \cite{fortunato2010community} for comprehensive surveys. These do not consider spatial interactions - leveraging these approaches in a geospatial context is an important possibility for future work. 

Geospatial networks have been explored with respect to problems other than community finding such as link-prediction~\cite{larusso12} and identifying user location~\cite{abrol12}.  There have also been several empricial studies on social networks with a spatial component such as \cite{barth11,cho11,eagle2006reality}.  More domain-specific empirical studies related to this work are also prevalent in the literature.  Pertinent to our application are studies on terrorist networks~\cite{medina} and criminal co-offender networks~\cite{Schaefer11}.

\section{Conclusion}
In this work, we introduced spatially-near modularity - a measure of the quality of a geographically-near partition in a social network.  Though finding an optimal partition with respect to this measure is NP-hard, we were able to obtain quality partitions with two heuristic algorithms that we introduced in this paper and tested on real-world datasets.  We have also discussed various ways in which our algorithms can be applied to gain useful knowledge in counter-terrorism applications. Our immediate concern for future work is exploring the scalability of this approach ($10^6$ nodes and greater). Additionally, we are also pursuing temporal dynamics of such communities and the differences between the communities formed based on the current state of the nodes (i.e. ``work'' vs ``home'').  In our more practical research, we are also working to integrate the generation of geographically-near partitions into our Organizational, Relationship, and Contact Analyzer (ORCA) software~\cite{paulo13fosint} that we are currently fielding to several American law-enforcement agencies.\\

\noindent\textbf{Acknowledgements.} This work was supported by ARO (project 2GDATXR042 and grant W911NF-08-1-0144), AFOSR (grant FA9550-12-1-0021) and the Office of the Secretary of Defense.  The opinions in this paper are those of the authors and do not necessarily reflect the opinions of the funders, the U.S. Military Academy, or the U.S. Army.

\end{document}